\documentclass{iaus}
\usepackage{graphicx}

\title[] 
{Modelling stellar activity cycles \\ using deep-seated dynamos and \\ surface 
flux transport}

\author{Emre I\c{s}\i k$^1$, Dieter Schmitt$^2$, \& Manfred Sch\"ussler$^2$}

\affiliation{$^1$Department of Physics, Istanbul K\"ult\"ur University,
Bak\i rk\"oy, 34156, Istanbul, Turkey \break email: e.isik@iku.edu.tr \\[\affilskip]
$^2$Max-Planck-Institut f\"ur Sonnensystemforschung, Max-Planck-Str. 2, 
37191 Katlenburg-Lindau, Germany}

\pubyear{2013}
\volume{294}  
\pagerange{119--126}
\jname{Solar and Astrophysical Dynamos and Magnetic Activity}
\editors{A.C. Editor, B.D. Editor \& C.E. Editor, eds.}
\begin{document}

\maketitle

\begin{abstract}
We investigate the relations between tachocline-based dynamos and the 
surface flux transport mechanisms in stars with outer 
convection zones. Using our combined models of 
flux generation and transport, we 
demonstrate the importance of the buoyant rise of 
magnetic flux, 
which physically determines the emergence latitudes and tilt angles 
of bipolar magnetic regions. The combined effects of 
the dynamo strength, flux rise, and surface transport lead to 
various cyclic and non-cyclic time series of total unsigned surface magnetic 
flux. 
\end{abstract}
The outer layers of G- and K-type stars exhibit turbulent convection, which couples 
with rotation to amplify, transport, and disperse magnetic fields. 
A combination of various effects of magnetic flux transport can lead to a diversity of 
magnetic behaviour in G and K stars with different rotation rates. 
We have recently applied our three-part composite model (\cite{isik07}) 
to investigate the effects of flux generation, rise, and transport in G-K main-sequence 
and evolved stars (\cite{isik11}), which we briefly summarise below.

In the reference model corresponding to the Sun, we used a solar convection 
zone (CZ) model of \cite{stix91} with an overshoot layer at the base. In this layer, we 
solved the dynamo equations numerically in 1D for both $\alpha$ and $\Omega$ 
effects operating co-spatially, with the nonlinear saturation level set by the buoyancy 
instabilities of toroidal flux tubes, which form in the same layer (\cite{ss89}). 
The resulting butterfly diagram for the toroidal field determines the 
probability density for flux tube eruption. Flux tubes on the limit of buoyancy 
instability then rise through the CZ, giving the emergence latitudes and tilt 
angles of bipolar regions at the surface. The emergent bipolar regions with solar 
area distribution are then transported by a surface flux transport model. 

We found that with increasing rotation rate the dynamo cycle period shortens, 
and the correlation between the deep-CZ overshoot dynamo and the surface 
emergence pattern is gradually lost. For cooler stars, deeper convection zones 
decrease this correlation even further. 
Moreover, surface flux transport effects are also 
important in determining cycle properties. For a Sun-like star with a rotation 
period of 9 days, we found non-cyclic surface activity, in spite of the cyclic 
dynamo in the interior. 

\begin{figure}
\begin{centering}
\includegraphics[width=.3\linewidth]{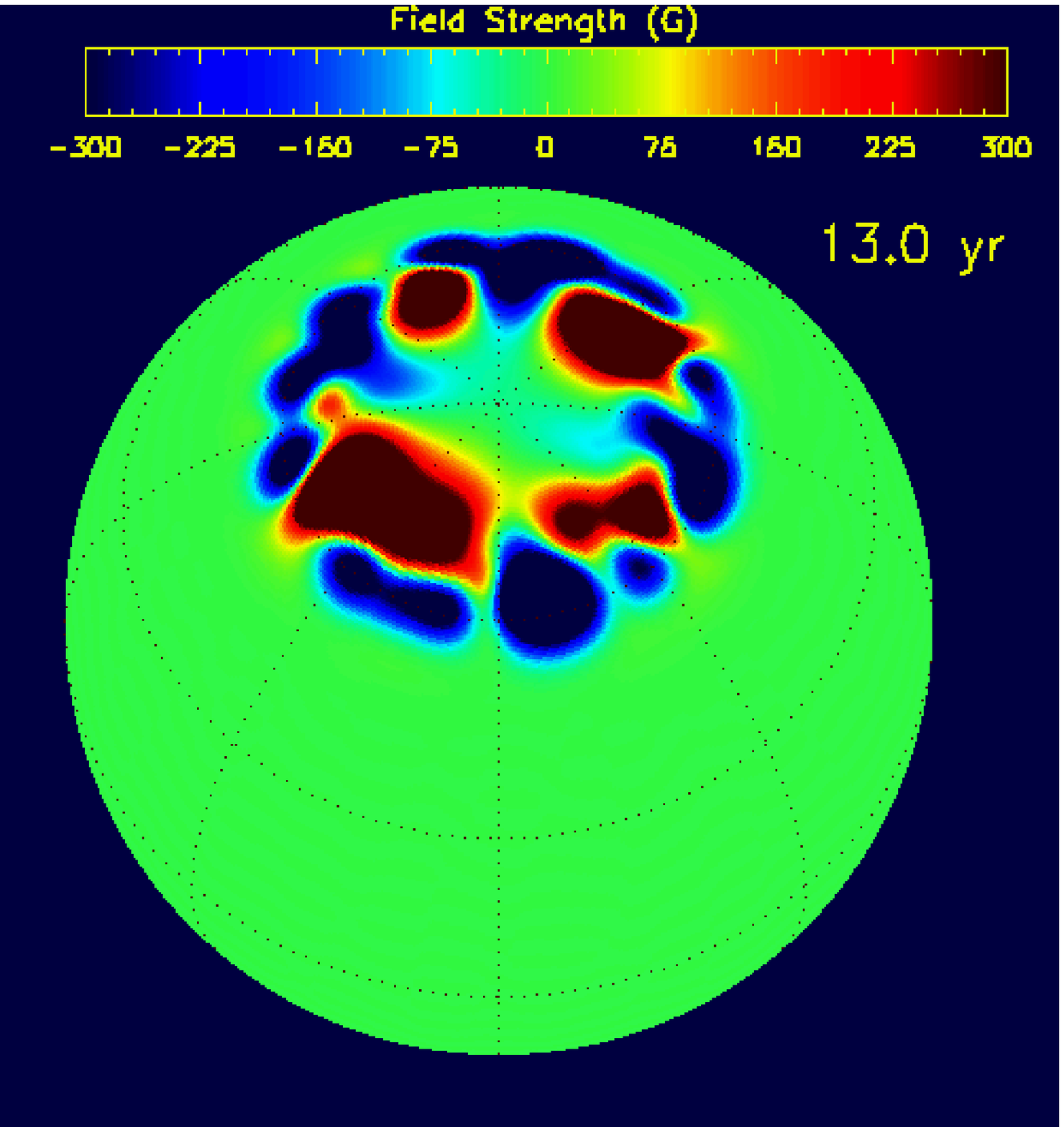}
\includegraphics[width=.3\linewidth]{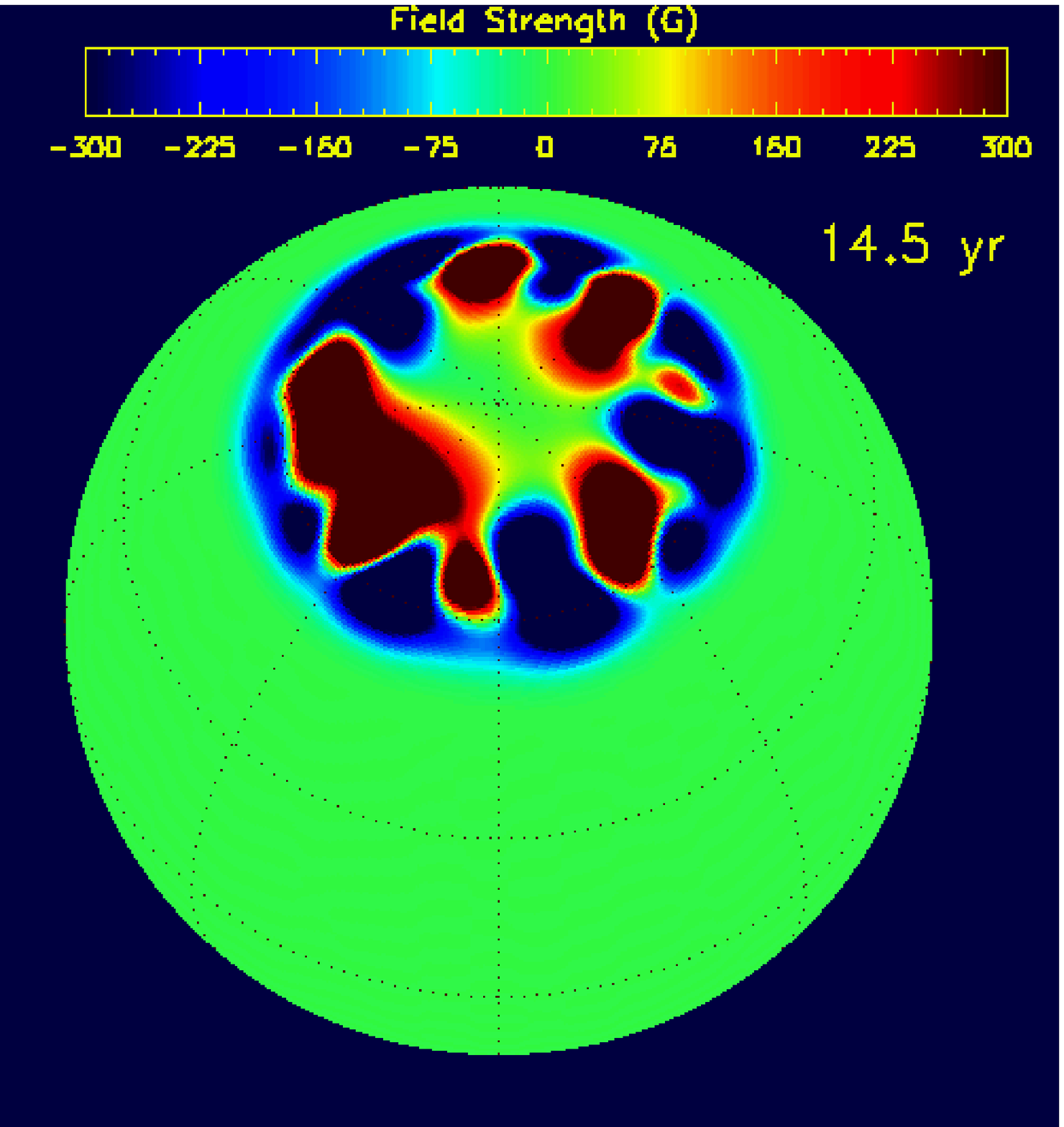}
\caption{Magnetic field distribution in a K1-type subgiant star similar to 
the active component of HR 1099, during activity minimum (left) and maximum 
(right).}
\label{fig:1}
\end{centering}
\end{figure}

\begin{figure}
\begin{centering}
\includegraphics[width=.8\linewidth]{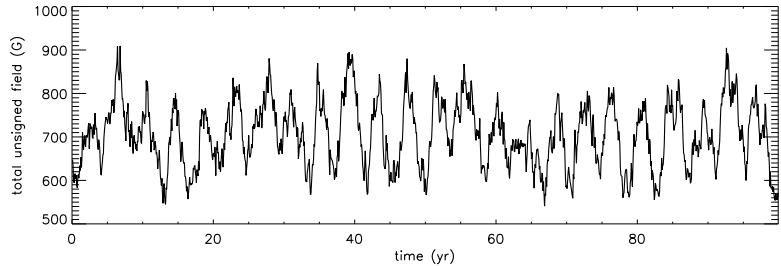}
\caption{Variation of the total unsigned flux density for the 
K1-type subgiant star in Fig.~\ref{fig:1}.}
\label{fig:2}
\end{centering}
\end{figure}

In the case of a K1-type subgiant star with the surface differential rotation, 
mass, and radius adopted from the K1IV component of HR 1099, we found 
that surface transport effects can lead to erratic variations in surface-integrated 
magnetic flux, which can be assumed to be correlated with stellar brightness 
(Figs.~\ref{fig:1}, \ref{fig:2}). 
In this model, a multi-periodic cycle pattern with periodicities 
similar to that of HR 1099 has been 
found (\cite{olah09}, \cite{isik12}). The mechanism responsible for the 
long-term variation in our model is that the stochastic emergence 
effects are of comparable importance with the surface transport effects and 
the dynamo-driven cyclic changes of polarity orientation of emerging bipolar 
regions. 

In the light of our numerical simulations, we underline the importance 
of non-local and nonlinear effects such as buoyant rise of flux tubes, 
particularly for lower Rossby numbers.
However, our models do not include the effect of the surface fields 
on the deep interior. To close the loop, our next attempt in stellar dynamo 
modelling is to construct a 
2D Babcock-Leighton type dynamo model including rising flux tube simulations 
in stars with different rotation rates and convection zone depths. 


\end{document}